# Quantum anomalous Hall effects controlled by chiral domain walls

Qirui Cui[1], Jinghua Liang[1], Yingmei Zhu[1], Xiong Yao[1], Hongxin Yang[1,2]*

[1]*Ningbo Institute of Materials Technology and Engineering, Chinese Academy of Sciences, Ningbo 315201, China*

[2]*Center of Materials Science and Optoelectronics Engineering, University of Chinese Academy of Sciences, Beijing 100049, China*

*Corresponding author: hongxin.yang@nimte.ac.cn

**Abstract**

We report the interplay between two different topological phases in condensed matter physics, the magnetic chiral domain wall (DW) and the quantum anomalous Hall (QAH) effect. We show that the chiral DW driven by Dzyaloshinskii–Moriya interaction (DMI) can divide the uniform domain into several zones where the neighboring zone possesses opposite quantized Hall conductance. The separated domain with a chiral edge state (CES) can be continuously modified by external magnetic field-induced domain expansion and thermal fluctuation, which gives rise to the reconfigurable QAH effect. More interestingly, we show that the position of CES can be tuned by spin current driven chiral DW motion. Several two-dimensional magnets with high Curie temperature and large topological band gaps are proposed for realizing these phenomena. Our work thus reveals the possibility of chiral DW controllable QAH effects.

The magnetic chiral DW is a type of topological defect with discrete symmetry, which is the boundary between domains with opposite magnetization and could be excited from uniform domain ground states. The topological charge of chiral DW is defined as: $Q_{DW} = (1/\pi) \int_{-\infty}^{\infty} \nabla\theta \, dx$, where $\theta$ represents the polar angle of normalized spin vector $S$. $Q_{DW}$ equals 1 or -1 when $S$ rotates from +z to -z or -z to +z. Domains separated by the chiral DW with topological protection is widely used as the information bit in emergent spintronic memory and logic devices, and the current-driven chiral DW displacement via spin-transfer torque (STT) or SOT underpins the operations of these devices [1-9]. Notably, one key term for stabilizing the chiral DW is DMI which favors the formation of noncollinear spin configuration in magnets lacking inversion symmetry [10-14].

QAH effect is another type of topological phases, which is characterized by the quantized Hall conductance $(Ce^2)/h$ without external magnetic field (where $C$, $e$, and $h$ represents Chern number, elementary charge, and Planck constant respectively). Due to its dissipationless CESs, QAH effect shows promising for applications in future electronic devices with ultralow-energy consumption [15-20] and provides an intriguing platform to investigate topological quantum physics, such as chiral topological superconductivity and Majorana fermions [21-25]. QAH effect is initially predicted by Haldane in 1988 [26] and first observed in magnetically doped topological insulator, Cr-doped (Bi, Sb)$_2$Te$_3$ thin films, by Xue et al in 2013 [17]. However, the extremely low full quantization temperature of 30 mK largely impedes its practical applications. Thus, tremendous efforts have been devoted to optimizing and designing material systems with high QAH effect temperature [27-33]. Besides high temperature, it is a long-sought goal for QAH effect that realizing effective manipulation of CESs, which probably leads to the artificial designing of quantum information transferring [34, 35]. In Cr-doped (Bi, Sb)$_2$Te$_3$ thin films, Yasuda et al demonstrate that two CESs would co-propagate along the DW [36-38] and first realize the reconfigurable CESs by using the tip of magnetic force microscope to write domain [34]. However, the investigation of interaction between two topological phases, chiral DW and QAH effect, in realistic materials remains very limited as far as we know, and particularly, it is still interesting and challenging to utilize chiral DW to control the high-temperature QAH effects.

As shown in Fig. 1, the coexistence of QAH effect and chiral DW require that materials combine nontrivial electronic states and sizable DMI. Importantly, the nontrivial topological gap should appear when magnetization is out-of-plane (OOP) and totally vanish when magnetization is in-plane (IP), thus resulting in chiral DW being an intrinsic boundary separating two parallel chiral states. Since CESs are intimately hinged with spin configurations, the approaches that are applied for controlling spin configurations will eventually lead to the CESs modification. For example, the spin vector can be aligned by a uniform magnetic field; spin fluctuations can be induced by laser or thermal excitations; and spin vector orientation can be explicitly and energy-efficiently controlled by spin current-generated torque. In the following, based on the first-principles calculations, Wannier-based tight binding models, and atomic spin model simulations, we first take VSe$_2$ monolayer with $P\bar{4}m2$ layer group as a representative example to demonstrate the manipulation of QAH effect via chiral DW, and then extend the discussions to other thin films, i.e., Fe$_2$XI (X=Cl, Br) Janus monolayers with $P4mm$ layer group. The computational methods are given in the supplemental materials (SM) [39].

The crystal structure of VSe$_2$ is shown in the Fig. S1(a)-(c). Each V atom is tetrahedrally surrounded by four Se atoms, and a Se atom bonding with two V atoms along *x* and *y* directions locates at bottom and top layer respectively. The calculations of phonon spectrum [Fig. S2(a)] demonstrates the dynamic stability. To investigate magnetic properties, we adopt the following spin Hamiltonian:

$$H = -J_1 \sum_{<i,j>} \mathbf{S}_i \mathbf{S}_j - J_2 \sum_{<i',j'>} \mathbf{S}_{i'} \mathbf{S}_{j'} - A \sum_i (S_i^z)^2 - \sum_{<i,j>} \mathbf{D}_{ij} \cdot (\mathbf{S}_i \times \mathbf{S}_j) - \mu B_{ext} \sum_i S_i^z, \quad (1)$$

where $J_1$ and $J_2$ represent the NN and NNN exchange coupling, respectively, *A* refers to the single-ion magnetic anisotropy, and $\mathbf{D}_{ij}$ refers to the DMI between NN V pairs. For extracting magnetic parameters in spin Hamiltonian, we apply the energy mapping methods [detailed discussion given in SM], and the results are shown in Table SI. $J_1$ in pristine VSe$_2$ reaches to 39.44 meV implying strong ferromagnetic exchange coupling between V atoms. Despite $J_2$ = -1.12 meV favoring antiferromagnetic coupling, its magnitude is much smaller compared with $J_1$. VSe$_2$ possesses perpendicular magnetic anisotropy of 0.47 meV which is an essential condition for achieving large size domain with OOP magnetization. Notably, due to the coexistence of $M_x$ and $M_y$ mirror symmetries, and $S_{4z}$ rotoreflection symmetry, the anisotropic DMI is allowed [orange arrows in Fig. S1(a)] according to the Moriya rules [11, 40]. Specifically, IP components of DMI between V pairs in $x$ ($d_\parallel^x$) and $y$ ($d_\parallel^y$) directions satisfy the relationship: $d_\parallel^x = -d_\parallel^y = 2.24$ meV, which is confirmed by the first-principles calculations [see Table SI].

We then perform atomic spin model simulations to investigate the spin textures since all magnetic parameters in spin Hamiltonian [Eq. (1)] are resolved. For describing the spin dynamics, the Landau-Lifshitz-Gilbert (LLG) equation is employed: $\frac{\partial S_i}{\partial t} = -\frac{\gamma}{(1+\alpha^2)}[S_i \times B_{eff}^i + \alpha S_i \times (S_i \times B_{eff}^i)]$, where $\gamma$ and $\alpha$ represents the gyromagnetic ratio and damping constant respectively. $B_{eff}^i$ indicates the effective field applied on each spin site and is defined as: $B_{eff}^i = -\frac{1}{\mu_s}\frac{\partial H}{\partial S_i}$. For pristine VSe$_2$, only uniform ferromagnetic state emerges arising from the strong ferromagnetic exchange coupling and perpendicular magnetic anisotropy, and Monte Carlo simulations confirm that Curie temperature reaches to 290 K [Fig. 2(a)]. For realizing noncollinear spin configurations, the amplitudes of $J$ and $A$ should be decreased while that of $d$ should be enhanced. These conditions are satisfied simultaneously by applying a slight tensile strain [Fig. S3]. Notably, single chiral DW with width of 6 nm and antiskyrmion solitons with diameter of 10 nm appear under 1% tensile strain [Fig. 2(b)-(d)]. For chiral DW, $Q_{DW}$ equals 1 as spin rotates from +z to -z, and for the observed antiskyrmion, topological charge $Q_{SK}$ is defined as $Q_{SK} = \frac{1}{4\pi}\int S \cdot (\partial_x S \times \partial_y S) dx dy$ and equals 1.

We now focus on the electronic states of VSe$_2$ monolayer. Fig. 3(a) shows the spin-polarized band structure. The spin-down channel exhibits a gap while spin-up channel is gapless at $\Gamma$ point. The projected band structures shows that electronic states nearby the Fermi level are dominated by the $p_x$ and $p_y$ orbitals of Se, and except for $\Gamma$ point, these two orbitals do not degenerate along -X $\leftrightarrow$ $\Gamma$ or $\Gamma$ $\leftrightarrow$ X lines in reciprocal space [Fig. S4]. When the spin-orbit coupling (SOC) effects are considered, a band gap of 116 meV are achieved accompanied with the $p_x$ - $p_y$ orbitals reversion [Fig. 3(b)], implying the emergence of topological electronic phase. Notably, if we apply the HSE06 rather than the standard GGA+U approach, the calculated band gap reaches to the 239 meV. For revealing topological properties, Wannier-based tight binding model is constructed by the $p_x$, $p_y$ orbitals of Se and $d$ orbitals of V. By integrating the Berry curvatures over the Brillouin zone, we obtain a quantized Berry phase of $\pi$ when the Fermi level locating in the band gap, which corresponds for the anomalous Hall conductivity $\sigma_{xy} = Ce^2/h$ with $C = 1$ [Fig. 3(c)]. The edge state calculations show that a gapless chiral edge mode appears in the band gap [Fig. 3(d)]. We thus demonstrate that VSe$_2$ is a QAH insulator with high critical temperature of 290 K. Notably, this QAH effect is robust to the external strain ranging from -5%-3% [Fig. 3(e) and (f)]. For demonstrating the boundary feature of chiral DW, we calculate the spin configurations-dependent band structures. The SOC-induced band gap appears in VSe$_2$ with OOP magnetization and vanishes when the magnetization is tuned to be IP [Fig. 3(g)]; and in a 50×1×1 supercell, the system with uniform ferromagnetism as ($\overbrace{\uparrow \cdots\cdots \uparrow}^{50\uparrow}$) clearly exhibits insulating properties, and interestingly, four edge modes emerge in band gap when two chiral DWs are introduced into spin configuration as ($\overbrace{\uparrow \cdots \uparrow}^{24\uparrow} \rightarrow \overbrace{\downarrow \cdots \downarrow}^{24\downarrow} \leftarrow$) [Fig. 3(h)].

Since chiral DW and QAH effect can coexist in VSe$_2$ under 1% tensile strain [Fig. 2(c) and 3(e)], we choose it as the example to reveal the chiral DW controllable QAH effects. As shown in the top panel in Fig. 4 ($B_{ext} = 0$ T), chiral DWs connect two edges of a strip and divides the uniform domain into several separate domains. The neighboring domain possesses opposite magnetization that opens the topological band gap with opposite Chern number ($C = +1$ or $-1$). Therefore, when the electrical current is injected from left-side current contact to right-side current contact, the emerging edge states of neighboring domain will exhibit the opposite chirality in the stripe [see lines with arrows in Fig. 4]. It is also expected that topological Hall effects are observed in the same system via slight electron/hole doping due to the existence of topological quasiparticles, antiskyrmion [41-43]. Via applying a non-zero $B_{ext}$, the domain with OOP magnetization expands accompanied with the chiral DW motion, and uniform ferromagnetic background is finally achieved as $B_{ext} = 0.4$ T. These results indicate that the zone of QAH state with $C = 1$ or $-1$ can be artificially designed via changing the magnitude of magnetic field. We further consider the temperature effects by assuming that thermal fluctuations of each spin are represented by a Gaussian white noise term. The thermal field on spin site is written as: $\boldsymbol{B}_{th}^i = \boldsymbol{\Gamma}(t)\sqrt{\frac{2\alpha k_B T}{\gamma \mu_s \Delta t}}$, where $\boldsymbol{\Gamma}(t)$ and $T$ represents the Gaussian distribution and temperature respectively, and the effective field is rewritten as: $\boldsymbol{B}_{eff}^i = -\frac{1}{\mu_s}\frac{\partial H}{\partial \boldsymbol{S}_i} + \boldsymbol{B}_{th}^i$. As shown in Fig. 4, the uniform ferromagnetic phase with single QAH state are destroyed by thermal fluctuation when temperature increases to 350 K and recover to the initial state with multiple QAH states by zero-field cooling. Thus, we achieve the reconfigurable QAH effects.

Chiral DW can also be driven by the spin current that is a lower energy-consumption and more convenient approach compared with external magnetic/temperature field. Usually, there are two distinct mechanisms to implement the spin current approach, i.e., STT and SOT. In STT, the spin-polarized current is injected into materials to transfer spin-angular momentum, which is impractical for the QAH insulator. In SOT, the spin accumulation at the interfaces exerts a torque on the magnetization of the adjacent magnet layer [see illustration of SOT in Fig. 1]. Interestingly, several vdW semiconductors, including $MoTe_2$, $WSe_2$ and $WTe_2$ etc., have been demonstrated to be able to generate sizable spin current via spin Hall effects or interface Rashba-Edelstein effects, and SOT-induced magnetization switching has been observed in these vdW semiconductors/ferromagnets heterostructures [44-48]. We include the SOT term $\boldsymbol{T}_{sot}^i$ in LLG equation for describing the spin dynamics of $VSe_2$ monolayer as: $\frac{\partial \boldsymbol{S}_i}{\partial t} = -\frac{\gamma}{(1+\alpha^2)}[\boldsymbol{S}_i \times \boldsymbol{B}_{eff}^i + \alpha \boldsymbol{S}_i \times (\boldsymbol{S}_i \times \boldsymbol{B}_{eff}^i) + \boldsymbol{T}_{sot}^i]$, where $\boldsymbol{T}_{sot}^i = \frac{\hbar J_c \theta_{sh}}{2e}\frac{a^2}{\mu_s}[\boldsymbol{S}_i \times (\boldsymbol{S}_i \times \boldsymbol{p}) - \alpha(\boldsymbol{S}_i \times \boldsymbol{p})]$. The first and second term of $\boldsymbol{T}_{sot}^i$ represents damping-like and filed-like torques, respectively. In atomic spin model simulations, $J_c$ is the current density; $\theta_{sh}$ is the spin Hall angle and is set to 0.1; and $\boldsymbol{p}$ is the orientation of spin polarization and is set to be along $+\boldsymbol{x}$. Fig. 5(a) shows the initial spin configurations and fully relaxed spin configurations after injecting 0.1 ns current with $J_c = 1\times10^{12}$ A/m². The chiral DW is driven by the SOT at an estimated high velocity $v$ of 623 m/s. We also find that $v$ linearly depends on $J_c$ and is enhanced to 805 m/s with $J_c = 1.4\times10^{12}$ A/m² [Fig. 5(b) and S5]. The motion of chiral DW naturally leads to the variation of CES positions, indicating that the accurate and fast controlling of a chosen zone of QAH state can be realized by adjusting the magnitude, polarization, or injecting time of spin current.

Finally, we show that coexistence of large-size domain separated by chiral DWs and high-temperature QAH effect can also be achieved in 2D Janus magnets, Fe$_2X$I ($X$=Cl, Br) monolayers, providing additional platforms for chiral DW controllable QAH effects. The inversion symmetry breaking of $P4mm$ layer group allows the DMI between Fe pairs. For resolving potential spin configurations, we adopt the following spin Hamiltonian:

$$H = -J_1 \sum_{<i,j>} S_i S_j - J_2 \sum_{<i',j'>} S_{i'} S_{j'} - A \sum_i (S_i^z)^2 - \sum_{<i,j>} D_{ij} \cdot (S_i \times S_j) - \sum_{<i',j'>} D_{i'j'}^{Cl/Br} \cdot (S_{i'} \times S_{j'}) - \sum_{<i',j'>} D_{i'j'}^{I} \cdot (S_{i'} \times S_{j'}). \quad (2)$$

Besides the NN DMI $D_{ij}$ [orange arrows of Fig. S6(a)], two additional terms, $D_{i'j'}^{Cl/Br}$ and $D_{i'j'}^{I}$, which are applied to describe the NNN DMIs between Fe pairs mediated by Cl/Br and I, respectively [blue arrows of Fig. S6(a)]. The magnitudes of IP components of above mentioned DMIs are labeled as $d_\parallel$, $d_\parallel^{Cl}$, $d_\parallel^{Br}$, and $d_\parallel^{I}$. As shown in the Table SII, sizeable DMI is achieved in both Fe$_2$ClI and Fe$_2$BrI, and notably, $d_\parallel^{I}$ of Fe$_2$ClI reaches 2.79 meV. The magnitudes of $d_\parallel$ and $d_\parallel^{I}$ are much larger than that of $d_\parallel^{Cl}/d_\parallel^{Br}$ due to the strong SOC scattering from I [12]. Atomic spin model simulation shows that uniform domain is divided into several regular zones by chiral DW [Fig. S6(b) and (c)]; and Monte Carlo simulations show that Curie temperature of Fe$_2X$I is around 400 K [Fig. S7]. We further discuss about the electronic states of Fe$_2X$I. When magnetization axis is tuned from $x$ to $z$, the SOC-induced Dirac gap of 211 and 253 meV is achieved for Fe$_2$ClI and Fe$_2$BrI respectively [Fig. S8(a)-(d)], and there are two gapless chiral edge modes emerging in band gap [Fig. S6(d) and (e)] confirming the emergence of QAH effects with high Chern number $C = 2$. It is thus expected that double CESs could be controlled simultaneously by chiral DWs in Fe$_2X$I, which is distinct from single CES in VSe$_2$ monolayer. We note that for Fe$_2X$I monolayers, coexistence of QAH effects and piezoelectricity has been demonstrated in Ref. 49, while the DMI, and it-favored topological spin configurations are revealed for the first time.

To summarize, we give a general approach of manipulation QAH effects and reveal its possibility in 2D magnets, VSe$_2$ and Fe$_2X$I (X=Cl, Br), with robust noncollinear magnetic order and high QAH temperatures. Specifically, the uniform domain in a strip is divided by DMI-favored chiral DWs into several domains where the neighboring possesses opposite quantized Hall conductance, and these separated domains with single/double CESs can be effectively tuned under the assistance of uniform magnetic and temperature fields, leading to the reconfigurable QAH effects. More interestingly, spin-orbit torque generated by spin current triggers translational motion of chiral DW at high velocity, which gives rise the precise and fast manipulation of QAH effect. Our findings thus open a previously unknown pathway to control the quantum transport of spin, which could benefit for novel and practical quantum applications.

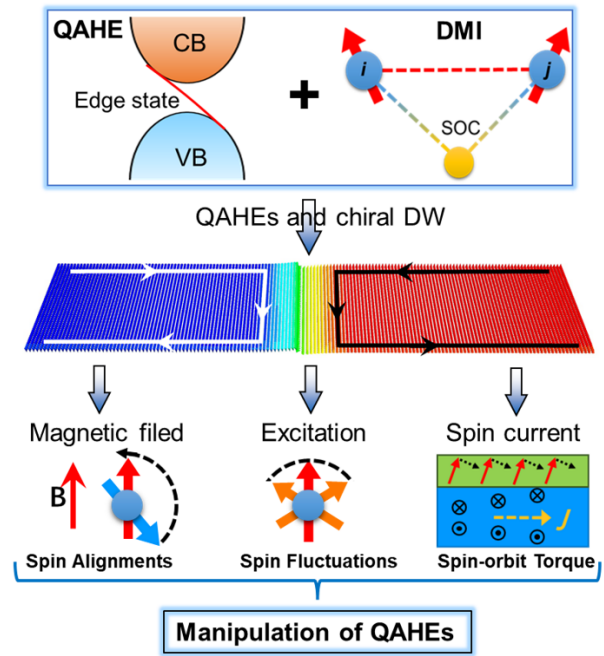

Fig. 1 The schematic of manipulation of QAH effects via chiral DW. For achieving the coexistence of these two topological phases, materials are required to combine nontrivial electronic states and sizable DMI. Due to CESs closely depending on the morphology of spin configurations, traditional approaches for tuning spin configurations all could lead to the CESs modifications.

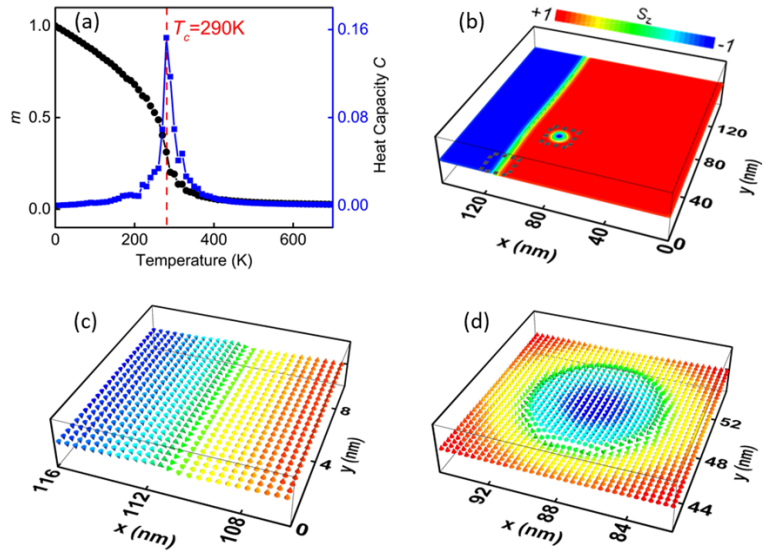

Fig. 2 (a) Normalized magnetization $m$ and heat capacity $C$ of pristine $VSe_2$ as the functions of temperature. (b) Spin configurations of $VSe_2$ with 1% tensile strain. The simulated zone is chosen to be a 150 nm × 150 nm square. (c) and (d) Zoom of spin textures of chiral DW and antiskyrmion as indicated by the black dashed in (b).

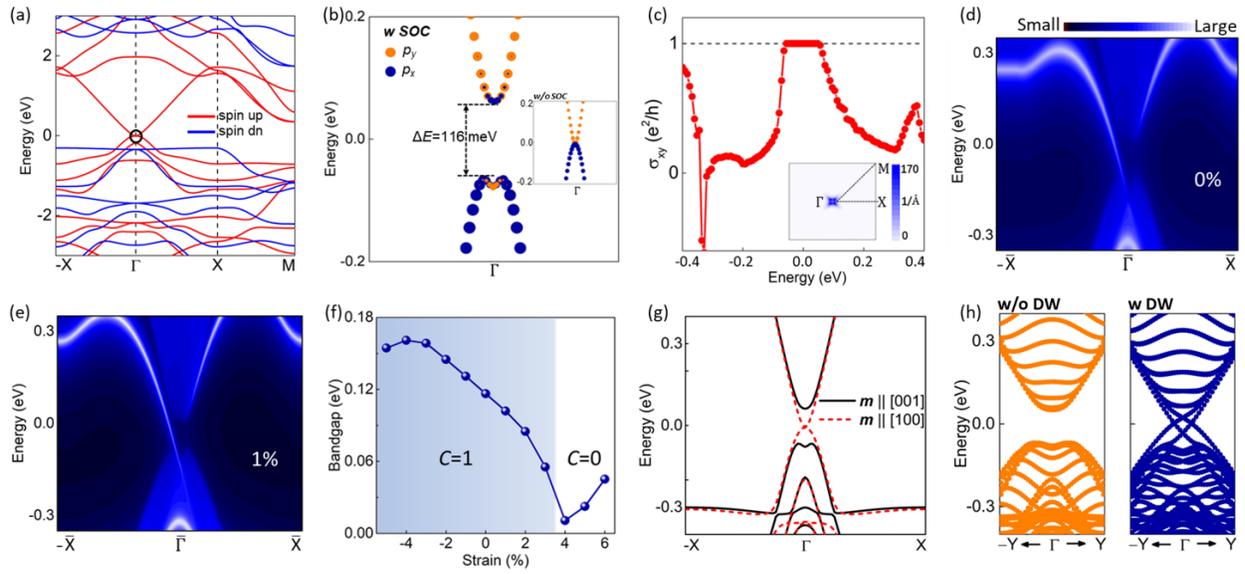

Fig. 3(a) Spin-polarized band structures of VSe$_2$. (b) Zoom of projected band structures around the Fermi level with considering the SOC effects. The inset shows band structures without considering SOC effects. (c) Anomalous Hall conductivity $\sigma_{xy}$ as the function of energy level. The inset shows the heat mapping of Berry curvature in Brillouin zone. (d) and (e) Edge states of semi-infinite plane of VSe$_2$ monolayer under 0% and 1% tensile strain. (f) SOC-induced bandgap as the function of strain. (g) Magnetization orientation-dependent band structures. (h) Band structures of a 50×1×1 supercell without and with considering chiral DW in spin configurations.

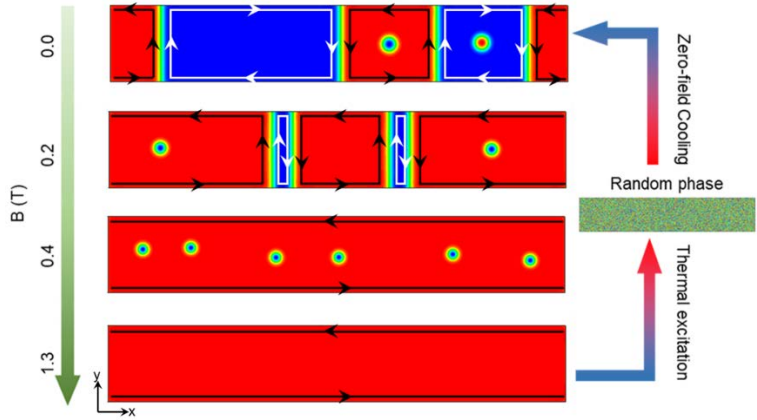

Fig. 4 (a) Reconfigurable QAH states controlled by external magnetic/temperature field. The top view of spin configurations of 300×50 nm VSe$_2$ stripe with open boundary. The black and white lines with arrows represent the quantized CESs in domain with up and down magnetization respectively.

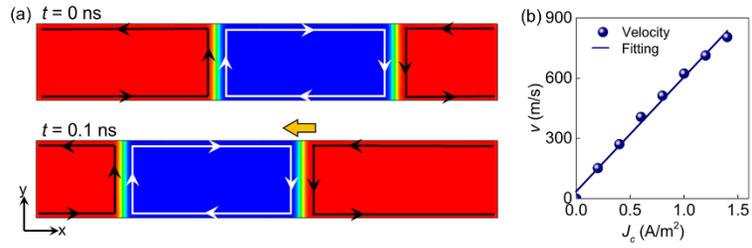

Fig. 5(a) The top view of VSe$_2$ stripe at initial state $t = 0$ ns and the later time stage $t = 0.1$ ns, under the time-invariant current density $J_c=10^{12}$ A/m$^2$. The orange arrow of down panel indicates the motion orientation of chiral DW. (b) Current density-dependent motion velocity of chiral DW.

**Acknowledgement**

This work was supported by the National Natural Science Foundation of China (Grant NOs. 11874059 and 12174405), Key Research Program of Frontier Sciences, CAS (Grant NO. ZDBS-LY-7021), Ningbo Key Scientific and Technological Project (Grant NO. 2021000215); "Pioneer" and "Leading Goose" R&D Program of Zhejiang Province (Grant NO. 2022C01053); Zhejiang Provincial Natural Science Foundation (Grant NO. LR19A040002).